\documentstyle[11pt,adassconf]{article}  % Leave intact

\begin{document}   % Leave intact

%-----------------------------------------------------------------------
%               Paper ID Code
%-----------------------------------------------------------------------
% Enter the proper paper identification code.  The ID code for your
% paper is the session number associated with your presentation as
% published in the official ADASS 2000 conference proceedings.  You can
% find this number locating your abstract in the printed proceedings
% that you received at the meeting or on-line via
% http://hea-www.harvard.edu/ADASS; the ID code
% is the letter-number sequence proceeding the title of your
% presentation.
%
% This will not appear in your paper; however, it allows different
% papers in the proceedings to cross-reference each other.  Note that
% you should only have one \paperID, and it should not include a
% trailing period.

\paperID{O02-3}

%-----------------------------------------------------------------------
%                   Paper Title
%-----------------------------------------------------------------------
% Enter the title of the paper.

\title{Hands-On Universe: A Global Program for Education
and Public Outreach in Astronomy}

\author{Michel Bo\"er, C. Thi\'ebaut}
\affil{Centre d'Etude Spatiale des Rayonnements (CESR/CNRS), BP
4648, F 31028 Toulouse Cedex 4 France}
\author{Hugues Pack}
\affil{Northfield Mount Hermon School, Northfield, Massachusetts,
USA}
\author{C. Pennypaker, M. Isaac}
\affil{University of California at Berkeley, USA}
\author{A.-L. Melchior}
\affil{DEMIRM/CNRS, Observatoire de Paris, France}
\author{S. Faye}
\affil{Lyc\'ee Jacques Decour, Paris, France}
\author{Toshikazu Ebisuzaki}
\affil{RIKEN, Tokyo, Japan}

\contact{Michel Boer}
\email{Michel.Boer@cesr.fr }

\paindex{Boer, M.} \aindex{Pack, H.} \aindex{Melchior, A.-L.}
\aindex{Pennypaker, C.} \aindex{Isaac, M.} \aindex{Ebisuzaki, T.}
\aindex{Faye, S.}


%-----------------------------------------------------------------------
%                  Abstract
%-----------------------------------------------------------------------
% Type abstract in the space below.  Consult the User Guide
% (http://hea-www.harvard.edu/ADASS)
% for a list of supported macros (e.g. for typesetting special
% symbols).

\begin{abstract}          % Leave intact
Hands-On Universe (HOU) is an educational program that enables
students to investigate the Universe while applying tools and
concepts from science, math, and technology. Using the Internet,
HOU participants around the world request observations from an
automated telescope, download images from a large image archive,
and analyze them with the aid of user-friendly image processing
software. This program is developing now in many countries,
including the USA, France, Germany, Sweden, Japan, Australia, and
others. A network of telescopes has been established among these
countries, many of them remotely operated, as shown in the
accompanying demo. Using this feature, students in the classroom
are able to make night observations during the day, using a
telescope placed in another country. An archive of images taken
on large telescopes is also accessible, as well as resources for
teachers. Students are also dealing with real research projects,
e.g. the search for asteroids, which resulted in the discovery of
a Kuiper Belt object by high-school students. Not only Hands-On
Universe gives the general public an access to professional
astronomy, but it is also a more general tool to demonstrate the
use of a complex automated system, the techniques of data
processing and automation. Last but not least, through the use of
telescopes located in many countries over the globe, a form of
powerful and genuine cooperation between teachers and children
from various countries is promoted, with a clear educational goal.

\end{abstract}

%-----------------------------------------------------------------------
%                 Main Body
%-----------------------------------------------------------------------
% Place the text for the main body of the paper here.  You should use
% the \section command to label the various sections; use of
% \subsection is optional.  Significant words in section titles should
% be capitalized.  Sections and subsections will be numbered
% automatically.

\section{Introduction}

The advantages of using Astronomy in the classroom are multiples:

\begin{itemize}

\item The Universe and its objects have to be learned
in most of the curricula, either as part of physics courses, or on
its own.

\item Astronomy may provide a good application of many of the
concepts developed in physics and mathematics curricula.

\item Astronomy is at the intersection of many areas of the
knowledge, either in fundamental and applied sciences. It is a
good illustration of the usefulness of an interdisciplinary
approach.

\item It is also a good illustration of the emergence of
science over the past centuries, and of the idea (still
developing) of the place of mankind in the Universe.

\item Many of the astronomical concepts have been developed in
various historical areas (Mesopotamia, Egypt, ancient Greece,
Arabic countries, Occidental world...).

\end{itemize}

 The goal of the Hands-On Universe (HOU) program is to promote the use
of astronomy within the high and middle schools and to enable students to
use data or to request there own observations from professional
or dedicated observatories. HOU has been historically developed
at the University of California Berkeley,
as a curriculum program. In this framework, a dedicated software
to analyse astronomical images has been written. Now, HOU has
been extended to more than nine countries over four continents,
and promotes the use of a global network of telescopes.

In this paper we present the main features of the HOU program,
and its global approach.

\section{Main teaching goals of HOU}

As explained above, beside Astronomy one of the main goals of HOU
is to use the concepts and data acquired in Astronomy to
introduce the scientific notions in physics and mathematics, for
high-school students. As an illustration the following concepts
may be introduced using astronomical data:

\begin{itemize}

\item In physics, the notion of speed of light may be illustrated
by reproducing the Romer's experience.

\item In mathematics, notions such coordinates, maps and
transformations may be illustrated using celestial reference
systems. Astronomical observation brings these transformations
from abstract to real life.

\item In image and signal processing, concepts like contrast may be
approached when pupils work with actual images.

\item An astronomical observatory has many complex instruments,
and if automated, it is a good illustration of automation for a
technology course (e.g. housekeeping).

\item Of course, an actual exploration of solar system bodies brings
"life" to an introduction to astronomy.

\end{itemize}

Though HOU has for main goal the teaching at the high and middle school
level, many demonstrations have been performed for the general
audience, e.g. at the open days of the University of Berkeley, at
the Science Museum of Tokyo, at the SITEF technological exhibition in
Toulouse, and at the Villette Science Museum in Paris.

A specific computer program has been developed which provides
access in a simplified form to most of the functions available in
astronomical data processing software. It has been
written for PC and Macintosh type computers, and give a clear interface for
pupils. The data processing functions include various color
tables, computation of sky level, aperture photometry, slice
extraction, histogram and axes plotting... Several of these
features are illustrated in figure 1 on an image of M 33 taken by
the TAROT telescope (Bo\"er et al. 2000; Bo\"er et al. 2001).

%\begin{figure}

%
%\epsscale{.70}

%\plotone{O02-3a.ps}
%\caption{The Messier 33 galaxy displayed using the HOU software,
%and featuring the "sky" and "slice" tools
%(image from TAROT, Bo\"er et al. 2001) }
%\label{O02-3a}

%\end{figure}

\section{Global HOU}

Global HOU (G-HOU) is an open network of teachers, high-schools, EPO and
research institutions, whose goal is the use of astronomy in
the various curricula. Several resources are shared, like
exercises, images, telescope time.

Several telescopes are giving time to the G-HOU network, e.g. the
TAROT instrument in France, the Keio observatory in Japan, the
Katzman observatory in the USA, etc. Some of them are
professional research instruments, while several others have been
developed first for educational purpose. Most of these telescopes have
capabilities to be remotely operated. This global network is
widely spread in longitude (and also in latitude) and has
for advantage that pupils may access telescopes during normal
classroom hours, while there is the night at the telescope site.
Several successful experiences have been made, e.g. during the
open days of the university of Berkeley, the SITEF exhibition
in France, or within the framework of the astronomy class of the
Tokyo Science Museum. They all bring the powerful (and somewhat
magic) possibility to see the night sky of other countries. This
feature, together with the use of archival, or pre-requested data,
open the possibility for students to perform their own scientific
program. As an example, for a study of the colors of stars, a
sequence of multi-color observations may be requested and the HR
diagram built directly from the data by the pupils. Another
striking example is the discovery of a Kuiper belt asteroid by the
astronomy class of Mount Northfield college
(\htmladdnormallink{http://www.nmhschool.org/astronomy} )).
This discovery has been confirmed by another high-school class,
showing the level of cooperation reached within the G-HOU program.
More generally speaking, G-HOU is a mean for children from various
countries and languages to exchange and collaborate.

There is however some difficulties in the fact that the
educational programs may be quite different between various
countries. For instance, there is no dedicated "astronomy" course in France
(and in most countries), while this happen quite often in the USA.
This means that astronomy has to be used as a tool to introduce
other concepts in physical sciences or mathematics, e. g.
distance, color and temperature... The use of astronomy, with "real
data" at hand has proven to be very effective to stimulate the
pupil interest in science courses, including in un-favorised
areas.

\section{Conclusion}

We presented the Hands-On Universe program, whose home page is
\htmladdnormallink{http://hou.lbl.gov/global/}. One of the most
interesting feature of this program is the use of actual, as well
as archived, data within the classroom and the ability for
students from various countries to build their own research
program and to work with "their" data.

Several exercises have been designed by the teachers participating
to the program, most of them aimed at introducing the notions and
concepts used in physical sciences, technology and mathematics
courses. Last but not least, through the use of
telescopes located in many countries over the globe, a form of
powerful and genuine cooperation between teacher and children
from various countries is promoted, with a clear educational goal.

%\acknowledgments

%-----------------------------------------------------------------------
%                 References
%-----------------------------------------------------------------------

%

% Do not place any material after the references section


\begin{references}
\reference Bo\"er M., et al., 1999, \aaps, 138, 579
 \reference Bo\"er et al. 2000, \adassix, \paperref{P2-19}
 \reference Bo\"er et al.\ 2001, \adassx, \paperref{D12}

\end{references}
\end{document}